\documentclass[prd,aps,showpacs,showkeys]{revtex4-1}
\usepackage{bm,amsfonts,latexsym,amsmath,amssymb,amsbsy,graphicx}
\newcommand{\bb}{\begin{eqnarray}}
\newcommand{\ee}{\end{eqnarray}}
\newcommand{\ba}{\begin{align}}
\newcommand{\ea}{\end{align}}

\begin{document}

\title{\bf Effect of vacuum polarization of charged massive fermions in an Aharonov--Bohm field}
\author{V.R. Khalilov}\email{khalilov@phys.msu.ru}
\affiliation{Faculty of Physics, M.V. Lomonosov Moscow State University, 119991,
Moscow, Russia}

\begin{abstract}
The effect of vacuum polarization of charged massive fermions in an
AharonovBohm (AB) potential in 2+1 dimensions is investigated.
The causal Green's function of the Dirac equation
with the AB potential is represented via the regular and
irregular solutions of the two-dimensional radial Dirac equation.
It is shown that the vacuum current density contains the contribution
from free filled states of the negative energy continuum as well as that
from a bound unfilled state, which can emerge in the above background
due to the interaction of the fermion spin magnetic moment
with the AB magnetic field while the induced charge density
contains only the contribution from the bound state.
The expressions for the vacuum charge and induced current densities are
obtained (recovered for massless fermions) for the graphene in the
field of infinitesimally thin solenoid perpendicular to the plane of
a sample. We also find the bound state energy as a function of magnetic flux,
fermion spin and the radius of solenoid as well as
discuss the role of the so-called self-adjoint extension parameter
and determine it in terms of the physics of the problem.
\end{abstract}

\pacs{03.65-w, 03.65.Ge, 03.65.Nk}

\keywords{Aharonov-Bohm potential; Bound state; Vacuum polarization; Vacuum charge; Vacuum current}

\maketitle


\section{Introduction}

Quantum systems of relativistic fermions in external fields in 2+1 dimensions attract considerable
interest related to the possibility of applying the results obtained for simplified models
to study fundamental physical phenomena such as the AharonovBohm and quantum Hall effects \cite{1,2}, as well as high-temperature superconductivity \cite{3}.
When the external field configuration has the cylindrical symmetry, a natural assumption is that the
relevant quantum mechanical system is invariant along the symmetry (z) axis and the system then
becomes essentially two-dimensional in the xy plane. Solutions of the Dirac equation with the
AB potential in 2+1 dimensions have been used to describe scattering of spin-polarized electrons by
infinitesimally thin solenoid  \cite{4} as well as the behavior of fermions in the field
of a cosmic string \cite{5}.

Important physical quantities are the vacuum charge and current
densities induced by the background field.
Vacuum polarization effects in the (2+1)-dimensional quantum electrodynamics
with homogeneous magnetic field and with a nonzero fermion density were studied
in \cite{angs,hvp}.

New interest to two-dimensional fermion systems with the energy spectrum
governed by Dirac Hamiltonian was revived in connection with problems of
 graphene (see, e.g., \cite{6} and \cite{7,8,9}).
In graphene, the electrons near the Fermi surface can be described
in terms of an effective Lorentz-invariant theory, with their kinetic energy
determined by Dirac's dispersion law, and the dynamics
of the electron at low energies is described by the Dirac equation
in 2+1 dimensions for a zero-mass fermion \cite{7}.
It should be noted that while a description of electron states in
 graphene in \cite{10,11,12} were based on the Dirac equation
for massless fermions, work \cite{13} has shown that the massive case
can also be created.

At the same time, the effective fine structure
constant in graphene is large, which leads to a new possibility
for studying quantum electrodynamics in the strong-coupling regime.
Charged impurity screening in graphene  in terms of vacuum polarization
 were investigated  in \cite{12,vp11a,as11b,kn11c,13a}.
 The  induced vacuum current in graphene
in the field of a solenoid was investigated in \cite{15} and a
wonderful phenomenon was revealed: the induced current turns out to be a
finite periodic function of the magnetic flux.
Induced vacuum condensates such as the induced charge density and current
(and other ones)
for massless fermions in the background
of a singular magnetic vortex in (2+1)-dimensional space-time
were investigated in \cite{31,32}.

The the Dirac equation with the AB potential is governed
by a (singular) Hamiltonian that requires
the supplementary definition in order for it to be treated
as a self-adjoint quantum-mechanical operator.
Based on the form asymmetry self-adjoint extension
method \cite{16,17}, the most relevant physical quantities,
such as energy spectrum, wave functions and the self-adjoint
extension parameter, were determined for some quantum systems with
the AB potential by applying the so-called self-adjoint boundary conditions
 in \cite{18,19,20,21}. It is useful to note that
the self-adjoint extension regularization were applied for two-dimensional
models in quantum field theory (see \cite{23,24}). As a result
some uncertainty appears in the prediction of physical
quantities \cite{23,24}. Essentially, a physical interpretation of
self-adjoint extensions is a purely physical problem and each extension
can (and must) be understood through an appropriate physical regularization \cite{17}.

In the present paper, we use the physical regularization
method for constructing the fermion wave function
and determining the energy in a bound state as a function of magnetic flux,
fermion spin. We introduce a small parameter
$R$, which, if we take in mind a real solenoid, is the
finite radius of this solenoid (see, for instance, \cite{4,25,26}.

We shall adopt the units where $c=\hbar=1$.

\section{Solutions to the Dirac equation in an Aharonov--Bohm field}

The space of particle quantum states in two spatial dimensions is the Hilbert space $\mathfrak H=L^2(\mathbb R^2)$ of square-integrable functions $\Psi({\bf r}), {\bf r}=(x,y)$  with the scalar product
\bb
(\Psi_1,\Psi_2)=\int \Psi_1^{\dagger}({\bf r})\Psi_2({\bf r})d{\bf r},\quad d{\bf r}=dxdy.
\label{scpr}
\ee

The Dirac equation for a fermion in a given external field can be obtained just as in 3+1 dimensions.
The Dirac $\gamma^{\mu}$-matrix algebra is known to be represented in terms of the two-dimensional Pauli matrices $\sigma_j$
\bb
 \gamma^0= \sigma_3,\quad \gamma^1=is\sigma_1,\quad \gamma^2=i\sigma_2 \label{1spin}
\ee
and the parameter $s=\pm 1$ can be introduced to label two types of fermions in accordance with the
signature of the two-dimensional Dirac matrices \cite{27} and is applied to characterize two states of the
fermion spin (spin "up" and "down") \cite{4}. Then, the Dirac Hamiltonian for a fermion of the mass
$m$ and charge $e=-e_0<0$ in an Aharonov--Bohm potential $A_0=0,\quad A_r=0,\quad A_{\varphi}=B/r,\quad r=\sqrt{x^2+y^2},\quad \varphi=\arctan(y/x)$ is
\bb
 H_D=\sigma_1P_2-s\sigma_2P_1+\sigma_3m,\label{diham}
\ee
where $P_\mu = -i\partial_{\mu} - eA_{\mu}$ is the generalized fermion momentum operator (a three-vector). The Hamiltonian (\ref{diham})
should be defined as a self-adjoint operator in the Hilbert space of square-integrable two-spinors
$\Psi({\bf r})$ with the scalar product (\ref{scpr}). The total angular momentum $J\equiv L_z+ s\sigma_3/2$, where
$L_z\equiv -i\partial/\partial\varphi$, commutes with $H_D$.

In real physical space, because of the existence of finite magnetic
flux inside solenoid $\Phi=2\pi B$ the term including the spin parameter appears
in the form of an additional delta-function interaction of spin with magnetic field of solenoid
\bb
{\bf H}=(0,\,0,\,H)=\nabla\times {\bf A}= 2\pi B\delta({\bf r})
\label{els}
\ee
in the Dirac equation squared. The additional potential
\bb
-seB\frac{\delta(r)}{r}
\label{els1}
\ee
will be taken into account by boundary conditions.
Such kind of point interaction also appears in several Aharonov--Bohm-like problems \cite{asp1,as0,sa1,fsa1}.

Eigenfunctions of the Hamiltonian (\ref{diham}) are (see, \cite{28})
\bb
 \Psi(t,{\bf r}) = \frac{1}{\sqrt{2\pi}}
\left( \begin{array}{c}
f_1(r)\\
f_2(r)e^{is\varphi}
\end{array}\right)\exp(-iEt+il\varphi)~, \label{three}
\ee
where $E$ is the fermion energy, $l$ is an integer. The wave function $\Psi(t,{\bf r})$  is an eigenfunction of the operator
$J$ with eigenvalue $j = l + s/2$. Taking into account the easily checked relations
\bb
sP_1\pm iP_2=-ie^{\pm is\varphi}\left[s\frac{\partial}{\partial r}\pm
\left(\frac{i}{r}\frac{\partial}{\partial
\varphi}-\frac{\mu}{r}\right)\right],\label{1rel}\ee
where $\mu\equiv e_0B$, we reduce the problem to that for the radial Hamiltonian $\check h$ in the Hilbert space of doublets $F(r)$ square-integrable on the half-line:
\bb
\check hF= EF, F=\left( \begin{array}{c}
f_1(r)\\
f_2(r)
\end{array}\right), \label{three}
\ee
where
\bb \check h\left(
\begin{array}{c}
f_1(r)\\
f_2(r)\end{array}\right)= \left[ \begin{array}{cc} m& sd/dr+(l+\mu)/r \\
-sd/dr+(l+\mu+s)/r& -m
\end{array}\right]\left( \begin{array}{c} f_1(r)\\
f_2(r)\end{array}\right) = E\left( \begin{array}{c} f_1(r)\\
f_2(r)\end{array}\right)\label{radh}\ee
in the range $r>0$.
Eliminating, for instance, $f_2(r)$ we derive the differential equation
for $f_1(r)$ and then  the lower (``small") component of doublet is found
from relation \bb
f_2(r)=-\frac{1}{E+m}\left(s\frac{df_1}{dr}-\frac{l+\mu}{r} f_1
\right). \label{lowc} \ee
Then, for $E^2-m^2>0$ the radial solutions  can be written via the Bessel functions:
\bb
F(r,E)=\left(
\begin{array}{c}
f_1\\
f_2
\end{array}\right)= A\left(
\begin{array}{c}
\sqrt{E+m}J_{\nu}(pr)\\
\pm \sqrt{E-m}J_{\nu\pm s}(pr)
\end{array}\right).\label{doubj}
\ee
Here $A$ is a constant, $\nu=|l+\mu|, p=\sqrt{E^2-m^2}$
and the upper (lower) signs should be taken for $l+\mu>0$
($l+\mu<0$).

Let us represent
\bb
\mu=[\mu]+\beta\equiv n+\beta,
\label{intfrac}
\ee
where $n$ denotes the largest integer $\leq \mu$, i.e. the integer part of $\mu$,
and $0<\beta<1$ is the fractional part of $\mu$. Hence $n = 0, 1, 2, . . .$
for $\mu>0$ and $n=-1,-2,-3, . . .$ for $\mu<0$.
Note that signs of $e$ and $B$ can be fixed and the  potential is
invariant under the changes  $e\to -e, s\to -s$, it hence suffices to consider only the case $e<0, \mu>0$. One can suppose that a bound state exists due to the interaction of the fermion spin magnetic
moment with AB magnetic field, which must be attractive. In the case  $\mu>0$, the potential
is attractive for $s=-1$ and repulsive for $s=1$, in the case $\mu<0$, it is attractive for $s=1$ and repulsive for $s=-1$. Then, it is seen that when $\beta\neq 0$ the upper and lower components of (\ref{doubj}) are integrable near $r=0$ only for $\nu\pm s>-1$.

Linearly independent at $\nu\neq 0$ solution $U(r;E)$ is determined as follows:
\bb
U(r,E)= C\left(
\begin{array}{c}
\sqrt{E+m}J_{-\nu}(pr)\\
\pm \sqrt{E-m}J_{-\nu\mp s}(pr)
\end{array}\right).\label{asdoub1}
\ee
Here $B$ is a constant. We shall need the irregular (i.e. integrable at $r\to \infty$)                                solutions for $E^2-m^2<0$. Such irregular solutions, nontrivial
at $\nu\neq n/2, n = 1, 2, . . .$ are the linear combination of
$F$ and $U$ and also can be represented via the MacDonald functions:
\bb
V(r,E)=\left(
\begin{array}{c}
v_1\\
v_2
\end{array}\right)
=B\left(
\begin{array}{c}
\sqrt{m+E}K_{\nu}(\lambda r)\\
\pm s\sqrt{m-E}K_{\nu\pm s}(\lambda r)
\end{array}\right), \label{doubm}
\ee
where $C$ is a constant and $\lambda=\sqrt{m^2-E^2}$.  The irregular solutions are
integrable at $r\to\infty$.

\section{A bound fermion state energy. Physical regularization}

A $\delta(x)$ potential is convenient to take into account artificially by means of the continuity relations. For this we replace the spin term (\ref{els1}) by (see, \cite{4})
\bb
-seB\frac{\delta(r-R)}{R}
\label{els2}
\ee
and take account of it by means of the continuity relations in $R$.
The quantity $B\delta(r-R)/R$ should not be considered as the real magnetic
field inside a flux tube but only as a field model allowing us to take
it into account by means of the continuity relations for solutions in
two ranges. Although the functional structure of Eqs. (\ref{els1})
and (\ref{els2}) are quite different, as discussed in \cite{26}, we are free
to use any form of potential provided that only the contribution
of the form (\ref{els1}) is excluded.

Now we can find the wave function and energy of bound state by means of
solutions to the Dirac equation in ranges $r<R$ and
$r>R$ and the potential (\ref{els2}) can be taken into account by means
of the continuity  relations. Obviously, for the model with the zero AB potential
in the range $r<R$ the radial solutions must satisfy (\ref{radh})
at $\mu=0$. They are written via the Bessel functions integrable near $r=0$:
\bb
S(r,E)=\left(
\begin{array}{c}
s_1\\
s_2
\end{array}\right)= C'\left(
\begin{array}{c}
\sqrt{E+m}J_{|l|}(pr)\\
\pm \sqrt{E-m}J_{|l|\pm s}(pr)
\end{array}\right), \label{freesol}
\ee
where $C'$ is a constant, $|l|\pm s\geq 0$ and the upper (lower) signs should be taken for $l>0$ ($l<0$).
The continuity relations can be written as
\bb
\left(
\begin{array}{c}
s_1\\
s_2
\end{array}\right)_{R-\delta}= \left(
\begin{array}{c}
v_1\\
v_2
\end{array}\right)_{R+\delta}, \quad \delta\to 0. \label{contin0}
\ee
The left- and right-hand sides of (\ref{contin0}) are calculated
using the asymptotic representation for the Bessel
functions in the limit $z\ll 1$:
\bb
 J_{\nu}(z)=\frac{z^{\nu}}{2^{\nu}\Gamma(1+\nu)}, \quad K_{\nu}(z)=-\frac{\pi}{2\sin(\pi\nu)}
\left[\frac{z^{\nu}}{2^{\nu}\Gamma(1+\nu)}-\frac{z^{-\nu}}{2^{-\nu}\Gamma(1-\nu)}\right].
\label{funcR}
\ee
Here $\Gamma(x)$ is the Euler gamma function of argument $x$.

The energy spectrum is determined by
\bb
\left(\frac{s_1}{s_2}\right)_{R-\delta}= \left(
\frac{v_1}{v_2}\right)_{R+\delta}, \quad \delta\to 0.
\label{spectr}
\ee
As a result, we obtain
\bb
 -s\sqrt{\frac{E+m}{m-E}}\frac{(kR)^{\pm s}\Gamma(1+|l|)}{2^{\pm s}\Gamma(1+|l|\pm s)}=
\left[\frac{(\lambda R)^{\nu\pm s}}{2^{\nu\pm s}\Gamma(1+\nu\pm s)}-\frac{(\lambda R)^{-(\nu\pm s)}}{2^{-(\nu\pm s)}\Gamma(1-(\nu\pm s))}\right]\phantom{mmmm}\nonumber\\
\left[\frac{(\lambda R)^{\nu}}{2^{\nu}\Gamma(1+\nu)}-\frac{(\lambda R)^{-\nu}}{2^{-\nu}\Gamma(1-\nu)}\right]^{-1}.\phantom{mmmmmmmmmmmmm}
\label{sew}\ee

It follows from Eq. (\ref{sew}) that the bound state energy is determined by the poles of gamma function $\Gamma(1+|l|\pm s)$ at $l=0$:
\bb
\frac{(\lambda R)^{-(n+\beta\pm s)}}{2^{-(n+\beta\pm s)}\Gamma(1-(n+\beta\pm s))}-\frac{(\lambda R)^{n+\beta\pm s}}{2^{n+\beta\pm s}\Gamma(1+n+\beta\pm s)}=1\pm s,
\label{eqpol}
\ee
where we must put $n=0$ and take the upper (lower) signs for $s=-1, \beta>0$ ($s=1, \beta<0$).
 In result, we
obtain the same equation for these two cases in the form:
\bb
\lambda=\frac{1}{R}\left(\frac{\Gamma(|\beta|)}{\Gamma(1-|\beta|)}\right)^{(2|\beta|-1)}.
\label{en1}
\ee

 We define particle (antiparticle) bound states as states that
tend to the upper (lower) continuous boundary  $m$ ($-m$) upon
adiabatically slow switching of the external field. Only
one particle (antiparticle) bound state with $s=-1$ ($s=1$) emerges.
 It is seen that these equations keep for the case $l+n=\mp 1$ ($s=\pm 1$). One can find from
(\ref{en1}) that an adiabatic increase of the magnetic flux
($\beta$) from $0$ to $1$ lifts a particle (antiparticle)
energy level $E=m\to E=-m$ ($E=-m\to E=m$) and for $|\beta|=1/2$
the particle and antiparticle energies are equal $E=\pm \sqrt{m^2-R^{-2}}$,
where the upper (lower) signs should be taken for particle (antiparticle).
We believe that doublet
\bb
V_0(r,E)
=N\left(
\begin{array}{c}
\sqrt{m+E}K_{\beta}(\lambda r)\\
 s\sqrt{m-E}K_{1-\beta}(\lambda r)
\end{array}\right) \label{wavegr}
\ee
($N$ is a normalization factor) represents  right
the particle radial wave function (with $s=-1$) in bound state.
It is evident, the wave function (\ref{wavegr}) is singular at $r=0$ but
square integrable on
the half-line $[0,\infty)$ with the measure $rdr$.

The bound state energy for the considered system was derived in \cite{21}
by means of the self-adjoint extension method  in the form:
\bb
\lambda=2m\left(-\frac{\Gamma(|\beta-1/2|+1/2)}{\xi \Gamma(1/2-|\beta-1/2|)}\right)^{-2(|\beta-1/2|)}.
\label{enss}
\ee
Here $\xi=\tan(\theta/2)$ ($2\pi\geq \theta\geq 0$) parameterizes
the self-adjoint extensions the radial Dirac
Hamiltonian, which are different for various $\theta$  except for two equivalent cases
$\theta=0, 2\pi$ (or $\xi=\pm \infty$).
We can determine the self-adjoint extension parameter in terms of the physics
of the problem, i.e. the parameter $R$.
By comparing equations (\ref{en1}) and (\ref{en1})  we arrive, for
example, for $\beta>1/2$
\bb
\xi=-(mR)^{2\beta-1}\frac{\Gamma(2-\beta)}{\Gamma(\beta)}.
\label{rxi}
\ee

\section{Vacuum charge and current densities}

Now we consider the densities of vacuum electric charge and
vacuum electric current due to  vacuum polarization. These quantities
are determined by the three-vector $j_{\mu}({\bf r})$, which is
expressed via the Green's function of the Dirac equation as follows
\bb
 j_{\mu}({\bf r})=-\frac{e}{2}\int\limits_{C}\frac{dE}{2\pi i}{\rm tr}G({\bf r}, {\bf r'}; E)\gamma_{\mu},
\label{cur0}
\ee
where $C$ is the path in the complex plane of $E$ enclosing all the singularities
along the real axis $E$ depending upon the choice of the Fermi surface
(we chose $E_F=-m$). The singularities of $G({\bf r}, {\bf r'}; E)$ can be
simple poles associated with the discrete spectrum (in the range $-m<E<m$),
and two cuts $(-\infty,-m]$ and $[m,\infty)$ associated with the continuum
spectrum in the ranges $|E|\geq m$. As was shown
in \cite{gr29}), for the partial Green's function in a Coulomb
field in 3+1 dimensions, the path $C$ may be
deformed to run along the singularities on the real $E$ axis as follows:
 $C=C_-+C_p+C_+$, where $C_-$ is the path along the negative
 real $E$ axis (${\rm Re}E<0$) from $-\infty$ to $0$ turning
 around at $E=0$ with positive
orientation, $C_p$ is a circle around the bound states' singularities with $-m< E<0$,
and $C_+$ is the path
along the positive real $E$ axis (${\rm Re}E>0$) from $\infty$ to $0$ but
with negative orientation (i.e. clockwise path)
turning around at $E=0$.
We note that due to the Furry theorem, the spatial component of induced vacuum current
in an AB potential in 2+1 dimensions should be an odd function of $\mu$. In the considered case
this is $j_{\varphi}$ - component of induced
vacuum current. It will be recalled that the $j_{\varphi}$ - component
is the vector product of vectors ${\bf j}$ and ${\bf n}={\bf r}/r$
and has the only component $[{\bf j\times n}]=j_xn_y-j_yn_x$;
it is the so-called pseudoscalar. Thus,
the $j_{\varphi}$-component of induced current is determined as follows
\bb
j_{\varphi}({\bf r})=-\frac{e}{2}\int\limits_{C}\frac{dE}{2\pi i}{\rm tr}G({\bf r},{\bf r'}; E)\gamma_{\varphi}.
\label{cur1}
\ee

For the Dirac equation in a Coulomb field in 3+1 dimensions, the radial partial Green's function is given by \cite{gr29}
\bb
G_l(r, r'; E)\gamma^0=\frac{1}{{\rm W}(E)}[\Theta(r'-r)\psi_R(r)\psi^{\dagger}_I(r')+
\Theta(r-r')\psi_I(r)\psi^{\dagger}_R(r')],
\label{green5}
\ee
where ${\rm W}(E)$ is the Wronskian and $\psi_R(r)$ and $\psi_I(r)$ are the regular and irregular
solutions of the radial Dirac equation $(\check H-E)\psi(r)=0$.
One can show that the Green's function in our case can be represented via the regular  and
irregular solutions of the two-dimensional radial Dirac equation just as in 3+1 dimensions.
It is convenient to apply doublet (\ref{doubj}) as the regular solutions and (\ref{doubm}) as the irregular ones. In such a way, we first construct the Green's function for the model with
the nonzero (at $r<R$) AB potential by analogy with the Coulomb case in 3+1 dimensions.
The ($r$-independent) Wronskian, defined by two doublets (\ref{doubj}) and (\ref{doubm})
as ${\rm Wr}(V, F)=rVi\sigma_2F=r(v_1f_2-f_1v_2)$, is easily calculated to be
\bb
{\rm Wr}(V, F)=\mp AC\lambda^{-\nu}p^{\nu}.
\label{wr1}
\ee
Here the upper (lower) signs should be taken for $l+\mu>0$ ($l+\mu<0$).

For the induced vacuum charge and current in the AB potential, we obtain
\bb
j_0(r)=-eN^2[K_{\beta}^2(x)+K_{1-\beta}^2(x)]_{\beta>1/2}-e\int\limits_{C_-+C_+}\frac{dE}{4\pi^2 i}{\rm tr}\left[\lambda^{\nu}p^{-\nu}\left((E+m)K_{\nu}(\lambda r)J_{\nu}(pr)- \right.\right. \nonumber\\ \left.\left. -s\sqrt{(m-E)(E-m)}\left(-K'_{\nu}(\lambda r) \pm s \frac{\nu}{\lambda r} K_{\nu}(\lambda r)\right)\left(\mp sJ'_{\nu}(pr)+\frac{\nu}{pr} J_{\nu}(pr)\right)\right)\right]
\label{vchar}
\ee
and
\bb
j_{\varphi}(r)=-2eN^2[K_{\beta}(x)K_{1-\beta}(x)]_{\beta>1/2}-e\int\limits_{C_-+C_+}\frac{dE}{4\pi^2 i}{\rm tr}\left[\lambda^{\nu}p^{-\nu}\left(pK_{\nu}\left(\mp sJ'_{\nu}(pr)+\frac{\nu}{pr} J_{\nu}(pr)\right)+\right.\right. \nonumber\\ \left.\left.
+s\lambda J_{\nu}(pr)\left(-K'_{\nu}(\lambda r) \pm s\frac{\nu}{\lambda r}K_{\nu}(\lambda r)\right)\right)\right].\phantom{mmmmm}
\label{vcur}
\ee
In (\ref{vchar}) and (\ref{vcur}) $N$ is the normalization factor,
the prime denotes the derivative of function with respect to argument
and ${\rm tr}\equiv \sum\limits_{l=-\infty}^{\infty}\sum\limits_{s=\pm 1}$; $x=\lambda(\beta, R)r$, where $\lambda(\beta, R)$ is determined by (\ref{en1}) so $x\sim c(r/R), c\sim 1$.

Using recurrent relations for the Bessel functions and summing over $s$, we find
\bb
j_0(r)=-eN^2[K_{\beta}^2(x)+K_{1-\beta}^2(x)]_{\beta>1/2} \pm e\int\limits_{C_-+C_+}\frac{dE}{4\pi^2 i}\sum\limits_{l=-\infty}^{\infty}\lambda^{\nu}p^{-\nu}(E+m)K_{\nu}(\lambda r)J_{\nu}(pr)
\label{vchar1}
\ee
and
\bb
j_{\varphi}(r)=-2eN^2[K_{\beta}(x)K_{1-\beta}(x)]_{\beta>1/2}-e\int\limits_{C_-+C_+}\frac{dE}{4\pi^2 i}\sum\limits_{l=-\infty}^{\infty}\lambda^{\nu}p^{-\nu}\frac{\nu}{r} K_{\nu}(\lambda r)J_{\nu}(pr).
\label{vcur1}
\ee
Integral over $E$ from the first term of integrand (\ref{vchar1}) gives $0$, because integrals taken along the paths $C_-$ and $C_+$ cancel in pairs. Now, it is possible to deform the paths $C_-$ and $C_+$ to the imaginary $E$ axis:
\bb
j_0(r)=-eN^2[K_{\beta}^2(x)+K_{1-\beta}^2(x)]_{\beta>1/2}\pm em\int\limits_{-\infty}^{\infty}\frac{dE}{4\pi^2}
\sum\limits_{l=-\infty}^{\infty}K_{\nu}(z)J_{\nu}(z)= j^b_0(r)+j^v_0(r)
\label{vchar2}
\ee
and
\bb
j_{\varphi}(r)=-2eN^2[K_{\beta}(x)K_{1-\beta}(x)]_{\beta>1/2}-\frac{e}{r}\int\limits_{-\infty}^{\infty}\frac{dE}{4\pi^2
}\sum\limits_{l=-\infty}^{\infty}\nu K_{\nu}(z)I_{\nu}(z)=j^b_{\varphi}(r)+j^v_{\varphi}(r).
\label{vcur2}
\ee
Here, symbols $b$ and $v$ characterize the contributions from bound and free states, respectively.
In (\ref{vchar2}) the upper (lower) signs should be taken for $l+\mu>0$ ($l+\mu<0$),
$I_{\nu}(z)=e^{-i\pi\nu/2}J_{\nu}(iz)$ is the modified Bessel function of first
kind of argument $z=\sqrt{m^2+E^2}r$.
An essential detail is that the vacuum charge ($j^b_0(r)$) and current ($j^b_{\varphi}(r)$) are singular at the origin, localized near the point $r=0$  and exponentially small
at $r\gg R$:
$$
j^b_0(r),\quad j^b_{\varphi}(r) \sim \frac{e}{r}e^{-cr/R}.
$$

Summation over $l$ in (\ref{vchar2}) gives $j_0^v(r)=0$.
For massless fermions, this result was obtained
and explained in \cite{15}
as follows: as far as  $j_0^v(r)$ should
be the odd function of $\mu$ due to the Furry theorem,
therefore, it must be pseudoscalar that would
contradict to the parity conservation of the two-dimensional
Dirac equation for massless particle. In
the two-dimensional model with mass term, this term is not
invariant with respect to the operations of time inversion and of spatial parity.
Nevertheless, there appears the nonzero vacuum charge density $j_0^b(r)$ due to
the vacuum charge of bound (empty!) state; the vacuum charge spatial distribution
is defined by the modulus squared of the fermion wave function
in the bound state (with $\beta>1/2$). The vacuum current density
contain the vacuum currents $j_{\varphi}^b(r)$ and $j_{\varphi}^v(r)$ that
are, respectively, due to the vacuum current of bound  state
and the vacuum current of free (filled!) states $j_{\varphi}^v(r)$,
which is spread out over an energy range of the negative energy continuum.
It is worth to note that the vacuum charge density
$j_0^v({\bf r})$ is induced by the homogeneous background magnetic field in
the massive ${\rm QED}_{2+1}$ \cite{angs,hvp}, but this is not so in the ${\rm QED}_{3+1}$.


Using the easily checked representation
\bb
K_{\nu}(z)I_{\nu}(z)=\int\limits_{0}^{\infty}dx e^{-2z\cosh x}I_{2\nu}(2z\sinh x)
\label{kirel}
\ee
and replacing variable $x$ by $y$ according to $\sinh x=\sinh^{-1}y$, we obtain
\bb
j^v_{\varphi}(r)=-\frac{e}{2\pi^2r}\int\limits_{0}^{\infty}dE \sum\limits_{l=-\infty}^{\infty}\nu \int\limits_{0}^{\infty}\frac{dy}{\sinh y} e^{-2z\coth y}I_{2\nu}(2z/\sinh y).
\label{mcurfin}
\ee
At $m=0$ (\ref{mcurfin}) takes the form
\bb
j^v_{\varphi}(r)=-\frac{e}{2\pi^2r}\int\limits_{0}^{\infty}dE \sum\limits_{l=-\infty}^{\infty}\nu \int\limits_{0}^{\infty}\frac{dy}{\sinh y} e^{-2Er\coth y}I_{2\nu}(2Er/\sinh y),
\label{0cur}
\ee
which coincides exactly with that for the induced vacuum current of massless fermions
for the first time  obtained in \cite{15}.

The integrals over $y$ from any summand of (\ref{0cur}) diverge, so
some quantity $\delta\ll 1$ should be introduced as a lower limit
of integration over $y$ in order to we could change the order
of summation and integration \cite{15}. After that we first take integral
over $E$ by means of formula \cite{GR}
\bb
\int\limits_{0}^{\infty}dt e^{-at}I_{\nu}(bt)=\frac{b^{\nu}}{\sqrt{a^2-b^2}(a+\sqrt{a^2-b^2})^{\nu}}.
\label{intt}
\ee
As a result, we obtain
\bb
j^v_{\varphi}(r)=-\frac{e}{2(\pi r)^2}\sum\limits_{l=-\infty}^{\infty} (l+n+\beta) \int\limits_{0}^{\infty}\frac{dy}{\sinh y} e^{-2|l+n+\beta|y}.
\label{1cur}
\ee
Taking the sum over $l$ by means of
\bb
4\sum\limits_{k=1}^{\infty} k e^{-2ky}=\sinh^{-2}y,
\quad 2\sum\limits_{k=1}^{\infty}e^{-2ky}=(e^y/\sinh y)-2
\label{sumk}
\ee
and then integral over $y$ with using formula \cite{GR}
\bb
\int\limits_{0}^{\infty}dt\frac{\sinh at}{\sinh bt} = \frac{\pi}{2b}\tan\frac{a\pi}{2b}, \quad b>|a|,
\label{finint}
\ee
we finally arrive at
\bb
j^v_{\varphi}(r)=\frac{e}{4\pi r^2}(2|\beta|-1)^2\tan\pi|\beta|.
\label{fincur}
\ee
We emphasize  that  $j^b_{\varphi}(r)=0$ at $m=0$ because the normalization constant $N\sim \sqrt{m}$. A massless charged fermion can not be bound with an AB potential.
As far as the main contribution to the integral over $E$
at $r\gg 1/m$ is given by the region $E\sim 1/r$, one can expect that
Eq. (\ref{fincur}) can be used for estimation of massive case if we  replace  $r^2$ by $r^2\sqrt{1+(mr)^2}$ in the denominator  of (\ref{fincur}).

One can see that  the induced current depends only on the fractional
part $\beta$ of $\mu$ and is finite periodical function of the magnetic flux.
For the first time these results were obtained in \cite{15}.

It should be noted that the nonzero probability current arises
under the effect of a constant uniform magnetic field on an electron
bound by an attractive delta-function potential \cite{29,30}
 The spatial distribution of probability current density resembles the spatial induced
 vacuum current $j^b_{\varphi}(r)$ but  the $j^b_{\varphi}(r)$ current arises due to
 vacuum polarization, i.e. when bound state is not filled.
We also note that vacuum polarization must manifest itself in
such a way so as to modify (change) the external potential.

 If we take in mind a real solenoid, then to learn the role
of finite small radius $R$ we consider again
the semirealistic model with the nonzero AB potential at $r<R$.
For such a model we can use the linear combination of
the $J_{\nu}(pr)$ and $J_{-\nu}(pr)$ Bessel functions
but not the Bessel ($J_{|l|\pm s}(pr)$) and
Neumann ($N_{|l|\pm s}(pr)$)  functions as
the  $\psi_R, \psi_I$ solutions for $r<R$. For $r>R$
the $K_{\nu}, I_{\nu}$ modified Bessel functions apply.
Then, continuity at $r=R$ determines the
linear combination of the $J_{\nu}$  and $J_{-\nu}$ solutions that joins
the $K_{\nu}$ function giving the irregular solution.  Continuity at $r=R$ also
can determine the particular linear combination of the $K_{\nu}$ and $I_{\nu}$
that joins the regular (i.e., integrable near $r=0$!).
With such constructed solutions the Greens function
will gain finite-size correction functions at $r<R$ and $r>R$ (see, \cite{gr29}).
It is wonderful that for the considered model we do not need to construct
the regular solutions in the range $r<R$ because we can apply regular solutions
constructed by means of the self-adjoint method in \cite{33}.
Regular solutions (doublets) must satisfy the so-called
self-adjoint boundary conditions [16]
\bb
(F^{\dagger}(r)i\sigma_2 F(r))|_{r=0}= (\bar f_1f_2-\bar f_2f_1)|_{r=0} =0. \label{bounsym1}
\ee
Physically, the self-adjoint boundary conditions show
that the radial component of probability current density is equal to zero
at the origin (the "origin" does not produce particles).
 For our problem the needed regular solution has the form
\bb
\psi(r)=\left(
\begin{array}{c}
f_1(pr)\\
f_2(pr)
\end{array}\right)
=D\left(\begin{array}{c}
\sqrt{E+m}[\cos(\theta/2)J_{(1+s)/2-s\beta}(pr)-\sin(\theta/2)J_{-(1+s)/2+s\beta}(pr)]\\
-s\sqrt{E-m}[\cos(\theta/2)J_{(s-1)/2-s\beta}(pr)+\sin(\theta/2)J_{(1-s)/2+s\beta}(pr)]
\end{array}\right),\label{irdoub}
\ee
where $D$ is the normalization factor and, as above, $1>\beta>0$, $\xi=\tan(\theta/2)$.
Note that the correct values of the self-adjoint parameter determine
the behavior of the upper (lower) component of doublet (\ref{irdoub}) at the origin.
Particularly, the case $\theta=0$ ($\theta=\pi$) is equivalent to insisting that
the upper (lower) component stays regular at the origin for any $s=\pm 1$ and, generally speaking, for $\mu>0$. If $\theta \neq 0, \pi$ both components of the doublet contain singular terms at the origin.

We need to estimate the contribution of the correction functions to the induced
vacuum current. To estimate the Green's correction function at $r\gg R$
we do not need to sum over $l$ and $s$, it hence suffices to consider
only the case, for example, $s=-1$. First,  from simple system
\bb
C_1K_{\beta}(z)+C_2I_{\beta}(z)= \cos(\theta/2)J_{\beta}(pR)-\sin(\theta/2)J_{-\beta}(pR)]\equiv f_1,\phantom{mmmmmmmmmm}\nonumber\\
C_1K_{1-\beta}(z)+C_2I_{1-\beta}(z)= \cos(\theta/2)J_{\beta-1}(pR)+\sin(\theta/2)J_{1-\beta}(pR)]\equiv f_2, \quad z=\sqrt{m^2+E^2}R, \phantom{mmmmm}
\label{est1}
\ee
we obtain
\bb
C_1=\frac{f_1I_{1-\beta}-f_2I_{\beta}}{W},\quad W=K_{\beta}I_{1-\beta}-K_{1-\beta}I_{\beta}.
\label{est2}
\ee

At $r\gg R$ the contribution of the Green's  (singular) correction
function to the integrand in Eq. (\ref{vcur1}) is determined with $C_1(ER)$.
Estimating $C_1$ with using Eq. (\ref{funcR}), one can
obtain: $C_1(ER)\sim (ER)^2\cos(\theta/2),  (ER)^{2\beta}\sin(\theta/2)$
The main contribution to the integral over $E$
at $r\gg R$ is given by $E\sim 1/r$ so that the contribution
of the Green's correction function to
the induced current is suppressed by the factors $(R/r)^2\cos(\theta/2),
(R/r)^{2\beta}\sin(\theta/2)$. These results are in agreement with estimations
made in \cite{15}.  The contribution of the Green's  (singular) correction
function to the vacuum charge density $j_0^v(r)$ may be nonzero and contain the same factors.

\section{Summary}

We have investigated the effect of vacuum polarization
for charged massive fermions in an AB
potential in 2+1 dimensions using the causal Green's function of
the Dirac equation with the AB potential represented via the regular and
irregular solutions of the two-dimensional radial Dirac equation,
which takes into account the fermion spin.
It is shown that the vacuum current density contains the contribution
from free filled states of the negative energy continuum as well as that
from a bound (empty!) state, which can emerge in the above background
due to the interaction of the fermion spin magnetic moment
with the AB magnetic field while the nonzero vacuum charge density
appears only due to the contribution from the bound state.
We have derived (recovered for massless fermions)  expressions for
the vacuum charge and vacuum current densities for graphene
in the field of infinitesimally thin solenoid perpendicular to the plane of
a sample.

A periodicity of the $j_{\varphi}^v$ vacuum electric current  due to to vacuum polarization
was observed recently in \cite{34a} in ``a quantum-tunneling system using two-dimensional ionic
structures in a linear Paul trap''. ``The charged quantum-tunneling particles should
be affected by the vector potential of a magnetic field throughout
the entire process, even during quantum tunneling,
and, thus, the AB effect should occur for tunneling particles''. Authors ``were successful
in observing the AB effect of tunneling particles using this system''.
It was revealed that ``the tunneling rate of the structure
periodically depends on the strength of the magnetic field, whose period is
the same as the magnetic-flux quantum $\phi_0$''. It will be noted that this result is in
agreement with  Eq. (\ref{fincur}), which is periodical function of the magnetic flux.

The work was supported in part by the Ministry of Education and Science of the Russian Federation
grant (Agreement No 14.576.21.0025 of 27.07.2014).


\begin{thebibliography}{55}

\bibitem{1} Y. Aharonov and D. Bohm, Phys. Rev., {\bf 115}, 485 (1959).
\bibitem{2} The Quantum Hall Effect, 4th ed. Editors: R.E. Prange, S.M. Girvin, (New York: Springer, 1990).
\bibitem{3} F. Wilczek, Fractional Statistics and Anyon Superconductivity, (World Scientific, Singapore, 1990).
\bibitem{4} C.R. Hagen, Phys. Rev. Lett., {\bf 64}, 503 (1990).
\bibitem{5} M.G. Alford and F. Wilczek, Phys. Rev. Lett., {\bf 62}, 1071 (1989).
\bibitem{angs} A.J. Niemi and G.W. Semenoff, Phys. Rev. Lett., {\bf 51}, 2077 (1983).
\bibitem{hvp} V.R. Khalilov, Theor. Math. Phys., {\bf 125}, 1413 (2000).
\bibitem{6} K.S. Novoselov et al., Science, {\bf 306}, 666 (2004).
\bibitem{7} A.H. Castro Neto et al., Rev. Mod. Phys., {\bf 81}, 109 (2009).
\bibitem{8} N.M.R. Peres, Rev. Mod. Phys., {\bf 82}, 2673 (2010).
\bibitem{9} V.N. Kotov et al., Rev. Mod. Phys., {\bf 84}, 1067 (2012).
\bibitem{10} K.S. Novoselov et al., Nature, {\bf 438}, 197 (2005).
\bibitem{11} Z. Jiang, Y. Zhang, H.L. Stormer, and P. Kim, Phys. Rev. Lett., {\bf 99} 106802 (2007).
\bibitem{12} I.S. Terekhov, A.I. Milstein, V.N. Kotov, and O.P. Sushkov, Phys. Rev. Lett., {\bf 100}, 076803 (2008).
\bibitem{13} F. Guinea, M.I. Katsnelson, and A.K. Geim, Nat. Phys., {\bf 6}, 30 (2009).
\bibitem{vp11a} V. M. Pereira, J.  Nilsson, and A.H. Castro Neto, Phys. Rev. Lett., {\bf 99}, 166802 (2007).
\bibitem{as11b} A.V. Shytov, M.I. Katsnelson, and L.S. Levitov, Phys. Rev. Lett., {\bf 99}, 236801 (2007).
\bibitem{kn11c} K. Nomura and A.H. MacDonald, Phys. Rev. Lett., {\bf 98}, 076602 (2007).
\bibitem{13a} I.F. Herbut, Phys. Rev. Lett., {\bf 104}, 066404 (2010).
\bibitem{15} R. Jackiw, A.I. Milstein, S.-Y. Pi, and I.S. Terekhov, Phys. Rev., {\bf B80}, 033413 (2009).
\bibitem{31} Yu. A. Sitenko, Phys. Rev., {\bf D60}, 125017 (1999).
\bibitem{32} Yu. A. Sitenko, Ann. Phys., 282, 167 (2000).
\bibitem{16} B.L. Voronov, D.M. Gitman, and I.V. Tyutin, Theor. Math. Phys., {\bf 150}, 34 (2007).
\bibitem{17} D.M. Gitman, I.V. Tyutin, and B.L. Voronov, Self-adjoint Extensions in Quantum Mechanics (Springer Science+Business Media, New York, 2012).
\bibitem{18} V.R. Khalilov and K.-E. Lee, Journ. Phys., {\bf A44}, 205303 (2011).
\bibitem{19} V.R. Khalilov, Theor. Math. Phys., {\bf 175}, 637 (2013).
\bibitem{20} V.R. Khalilov, Eur. Phys. J., {\bf C73}, 2548 (2013).
\bibitem{21} V.R. Khalilov, Eur. Phys. J., {\bf C74}, 2708 (2014).
\bibitem{23} Ph. Gerbert, Phys. Rev., D40, 1346 (1989).
\bibitem{24} M.G. Alford, J. March-Pussel and F. Wilczek, Nucl.Phys., {\bf B328}, 140 (1989).
\bibitem{25} V.R. Khalilov and C.-L. Ho, Ann. Phys., {\bf 323}, 1280 (2008).
\bibitem{26} E.O. Silva, F.M. Andrade, C. Filgueiras, and H. Belich, Eur. Phys. J., {\bf C73} (4) 2402 (2013).
\bibitem{27} Y. Hosotani, Phys. Lett., {\bf B319}, 332 (1993).
\bibitem{asp1} F.M. Andrade, E.O. Silva, M. Pereira, Phys. Rev., {\bf D85}(4), 041701(R) (2012).
\bibitem{as0} F.M. Andrade, E.O. Silva, Phys. Lett., {\bf B719}(4-5), 467 (2013).
\bibitem{sa1} E.O. Silva, F.M. Andrade, Europhys. Lett., {\bf 101}(5), 51005 (2013).
\bibitem{fsa1} C. Filgueiras, E.O. Silva, F.M. Andrade, J. Math. Phys.,
{\bf 53}(12), 122106 (2012).
\bibitem{28} V.R. Khalilov and K.-E. Lee, Mod. Phys. Lett., {\bf A26}, No 12, 865 (2011).
\bibitem{gr29} W. Greiner, J. Reinhardt, {\sl Quantum Electrodynamics},
$4^{th}$ ed. (Springer-Verlag, Berlin Heidelberg, 2009).
\bibitem{GR} I.S. Gradshteyn and I.M. Ryzhik, {\sl Table of Integrals,  Series, and Products}, $5^{th}$ ed. (Academic Press, San Diego, 1994).
\bibitem{29} F.Kh. Chibirova and V.R. Khalilov, Mod. Phys. Lett., {\bf A20}, 663 (2005).
\bibitem{30} V.R. Khalilov and F.Kh. Chibirova, J. Phys. A: Math. Theor., {\bf 40}, 6469 (2007).
\bibitem{33} V.R. Khalilov, K.-E. Lee, and I.V. Mamsurov, Mod. Phys. Lett., A27, No 5, 1250027 (2012).
\bibitem{34a}  A. Noguchi, Y. Shikano, Kenji Toyoda, Shinji Urabe, Nature Communications {\bf 5}, 3868 (2014).









\end{thebibliography}
\end{document}